# Three-Dimensional All-Dielectric Photonic Topological Insulator

Alexey Slobozhanyuk[1,2], S. Hossein Mousavi[3], Xiang Ni[1,4], Daria Smirnova[1,2], Yuri S. Kivshar[2], and Alexander B. Khanikaev[1,4]

[1]Department of Physics, Queens College of the City University of New York, Queens, NY 11367, USA

[2]Nonlinear Physics Centre, Australian National University, Canberra ACT 0200, Australia

[3]Microelectronics Research Centre, Cockrell School of Engineering, University of Texas at Austin, Austin, TX 78758 USA

[4]Graduate Center of the City University of New York, New York, NY 10016, USA

E-mail:akhanikaev@qc.cuny.edu

**The discovery of two-dimensional topological photonic systems has transformed our views on electromagnetic propagation and scattering of classical waves, and a quest for similar states in three dimensions, known to exist in condensed matter systems, has been put forward. Here we demonstrate that symmetry protected three-dimensional topological states can be engineered in an all-dielectric platform with the electromagnetic duality between electric and magnetic fields ensured by the structure design. Magneto-electric coupling playing the role of a synthetic gauge field leads to a topological transition to an "insulating" regime with a complete three-dimensional photonic bandgap. An emergence of surface states with conical Dirac dispersion and spin-locking is unimpeded. Robust propagation of surface states along two-dimensional domain walls defined by the reversal of magneto-electric coupling is confirmed numerically by first principle studies. It is shown that the proposed system represents a table-top platform for emulating relativistic physics of massive Dirac fermions and the surface states revealed can be interpreted as Jackiw-Rebbi states confined to the interface between two domains with opposite particle masses.**

Following the footsteps of condensed matter topological systems, a significant progress has been recently made in understanding and realizing topological states for bosons [1, 2, 3, 4, 5] and in classical systems [6, 7, 8, 9, 10, 11, 12, 13, 14 15, 16, 17, 18, 19, 20, 21, 22, 23]. Unlike fermionic systems, achieving topological order for bosons meets several limitations. In particular, while fermions can support robust topological phases protected by time-reversal (TR) symmetry alone, in bosonic and classical systems TR symmetry is not sufficient to protect any nontrivial topological phases [24]. Consequently, the traditional approach to engineering topological order in classical systems relies on removal of TR symmetry. This approach requires either using magnetic materials, e.g. in magnetic-photonic crystals emulating Quantum Hall effect (QHE) which were successfully realized in 2D [9, 25] and has also been recently extended theoretically to three dimensions [26], or temporal modulation emulating the effect of external magnetic field [27, 28, 29, 17, 30, 13].

Recently, a new class of topological states of condensed matter characterized by symmetry protected topological (SPT) phases has expanded the traditional topological classification based on time-reversal (TR) symmetry. As opposed to the TR symmetric systems, topological properties of SPT class originate in crystalline or intrinsic symmetries of the wave-fields. In the context of classical waves, the SPT phases represent attractive means to engineer topological order without need to break TR symmetry [15, 21, 31, 32], which can be a very restrictive requirement for electromagnetic and acoustic systems. Indeed, while in acoustics TR symmetry can be broken by magneto-elastic interactions; these are, however, are very small to be of any practical interest for realizing topological order. The same limitation applies to electromagnetics in optical domain, where magneto-optical effects are not strong enough to yield the topological order. However, even in microwave domain where magneto-optical effects can be very strong, the necessity to deal with magnets and magnetic materials, which are hard to integrate into practical devices, makes SPT phases more appealing for realizing topological order.

SPT phases have indeed been shown to result in a "symmetry restricted" topological order, where a conserved pseudo-spin degree of freedom can be judiciously engineered based on either spatial symmetries [33] or polarization degrees of freedom of a vector field [15, 21]. This approach has been especially fruitful in electromagnetics [24] where numerous theoretical predictions and experimental demonstrations have been made from microwave to optical frequencies [34, 35, 12, 15, 14, 36, 37, 23]. Although the existence of 2D topological states of light with TR symmetry is now a matter of experimentally confirmed fact, three-dimensional (3D) topological systems with preserved time-reversal symmetry has so far evaded emulation in photonics. And although time-reversal violating SPT topological insulator with a single Dirac cone has been recently put forward [26], the design suggested is rather challenging to implement in practice not only because of ferrites used in the design, but also due to the requirement of non-uniform magnetization of magnetic constituents in the proposed structure. Nonetheless, the possibility to engineer the topological order for photons in three dimensions and relying on a fabrication friendly platform is of a significant fundamental interest as it may allow emulating exotic quantum states of matter and relativistic quantum systems with Dirac dispersion. Moreover, it also may usher in a broad range of practical applications enabling topologically robust routing of photons in three dimensions, which envisions novel photonic technologies based on photonic elements integrated into 3D topologically robust optical circuitry.

In the following, we demonstrate a 3D topological photonic metacrystal based on the all-dielectric metamaterial platform [38, 39, 40, 41], which is thus free of very restrictive requirement of earlier concepts, including time-reversal violating systems, where magnets or temporal modulation are required, and approaches utilizing lossy metal-based metamaterial components [15, 31]. The suggested design can therefore be readily implemented across entire electromagnetic spectrum from microwave to visible domain.

In what follows we show by rigorous numerical simulations that SPT nontrivial states can be engineered in metacrystals preserving electromagnetic duality – the internal symmetry of the electromagnetic field responsible for the SPT phase [ 15, 31, 42, 23]. To this end, the duality of the electromagnetic eigenmodes, otherwise broken by materials response [ 15], is restored by careful design of building blocks of the three-dimensional photonic lattice referred to as meta-atoms. This ensures the presence of a proper pseudo-spin degree of freedom which is odd under TR operation and enables emulations of weak 3D topological order for light. An introduction of bianisotropy [ 43] into such dual systems is brought by proper reduction of the meta-atoms symmetry [ 15, 31, 23, 44] and induces an effective spin-orbital interaction [ 15, 45], giving rise to a topological transition accompanied by opening of a complete 3D photonic band-gap. A direct mapping of the 3D photonic topological insulator onto its condensed matter counterpart is demonstrated using two complimentary analytical approaches i) an effective Hamiltonian approach based on electromagnetic perturbation theory (Supplement A) and ii) an approach based on effective medium response (Supplement B). We show, both analytically and with the use of full-wave numerical calculations, that the two-dimensional domain walls, which are defined by the reversal of bianisotropy across all possible cut-planes of the 3D metacrystal, support surface states that exhibit Dirac-like conical dispersion. We demonstrate numerically that the surface states supported by topological domain walls are robust and avoid backscattering and localization.

The topological photonic metacrystal introduced here is schematically shown in Fig. 1(a). It represents a 3D hexagonal lattice of dielectric discs with the permittivity $\epsilon_d$ embedded into a matrix of permittivity $\epsilon_b$. The dielectric disks (meta-atoms) and their 3D periodic arrangement are designed in a way that the photonic band structure (PBS) exhibits two overlaid 3D Dirac points near the K- and K′-point in the Brillouin zone shown in Fig. 1. The structure parameters were optimized so there is a frequency range where only 3D Dirac cones exists, i.e. there are no any other modes for any Bloch vector within the 3D Brillouin zone that would belong to this range. Each of the four observed Dirac bands originates from the electric and magnetic dipolar modes of the discs with their dipole moments aligned in the $xy$ plane, perpendicular to the axes of the discs, as shown in Fig.2(a). It is important that the very possibility to emulate a true 3D Dirac dispersion with the four linear bands overlaid in a pairwise manner, the characteristic of relativistic massless spin-½ fermions described by the Dirac equation, is enabled by the duality symmetry of the system. Electromagnetic duality is a natural property of the electromagnetic field in free space which reflects symmetry of nature with respect to the electric and magnetic components of electromagnetic waves [ 46]. This symmetry, which is typically broken by uneven materials response to the electric and magnetic fields, is restored in our structure for the in-plane ($xy$) components by its careful design, and, therefore, is accidental and limited to a particular frequency range of interest. The presence of these two independent sets, the magnetic-dipole and electric-dipole degrees of freedom (essentially in-plane electric and magnetic p-orbitals), is ensured by the inversion symmetry in the z-direction, as in the case of metacrystal Fig.1 (a) [ 15, 21], as the electric and magnetic modes have opposite parity. Electric modes are even and magnetic modes are odd with respect to the $\sigma_z$ ($z \to -z$) transformation. Any reduction of the $\sigma_z$ symmetry in our structure results in the coupling of the in-plane electric and magnetic orbitals and opening of a band gap in the place of Dirac points. This magneto-electric coupling, referred to as bianisotropy, has an effect equivalent to the spin-orbital interaction in condensed matter systems (Supplements A and B) [ 15, 31, 21].

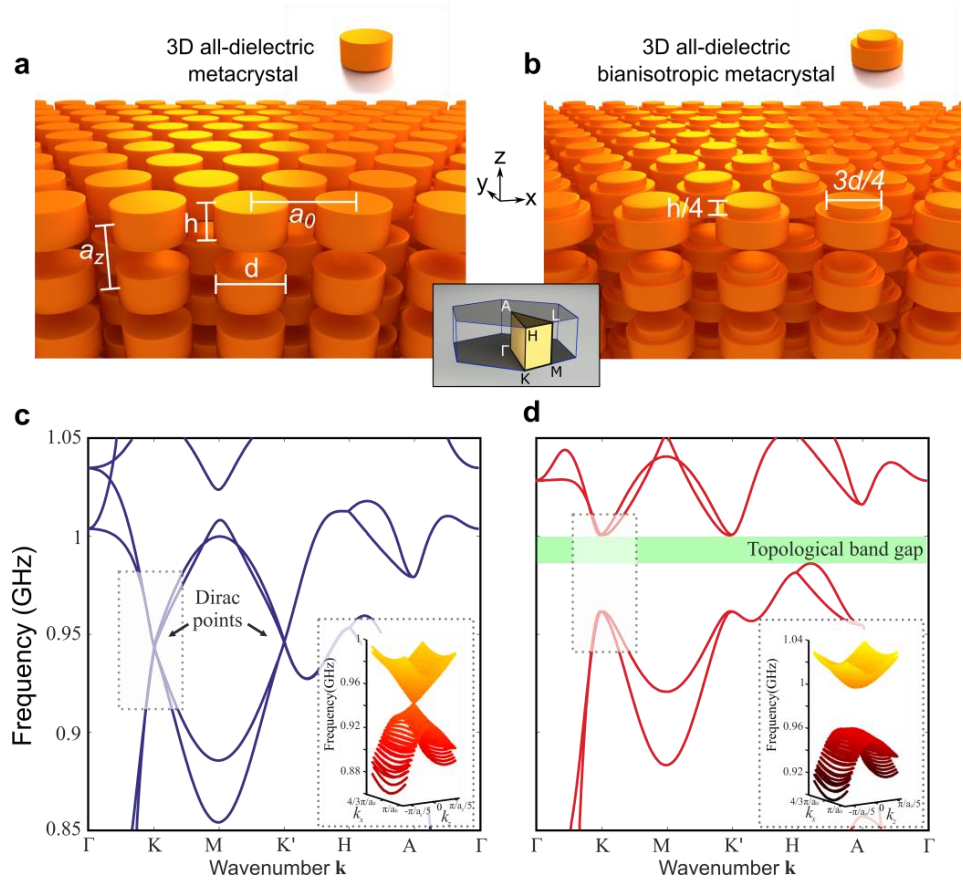

**Figure 1| Three-dimensional Dirac cone and topological transition in all-dielectric metacrystals.** Schematics of the metacrystals without (a) and with (b) bianisotropy and corresponding photonic band structures (c) and (d), respectively, calculated along the high symmetry directions. Brillouin zone of the 3D hexagonal lattice is shown as an inset between (a) and (b). The insets in (c, d): 2D projection of the 3D Dirac cone in the vicinity of K-point and emergence of the topological band gap induced by the bianisotropy. The structure parameters are $a_0 = 4.76$ cm, $a_0 = 4.49$ cm, $d = 4.76$ cm, $h = 2.11$ cm, $\epsilon_d = 81$, $\epsilon_b = 3$.

An example of the metacrystal with $\sigma_z$ symmetry broken by removing part of the dielectric cylinder is shown in Fig. 1(b), right panel, with its band structure in Fig. 1(d), and clearly reveals a full (omnidirectional) 3D bandgap. The modes of such gapped structure are no longer pure electric and magnetic modes, but represent mixed states with the phases between their electric and magnetic components being locked [15, 31]. The dipolar electric and magnetic moments of these modes always appear to be tilted by 90 deg and -90 deg with respect to each other, thus constituting two sets of eigenmodes $\psi^\uparrow$ and $\psi^\downarrow$, where the superscript indicates up (↑) or down (↓) pseudo-spin value, both for lower and upper bands.

As follows from the two complimentary techniques used, an analytical effective medium theory and electromagnetic perturbation theory based on numerical simulations, near the K (and K′) point the metacrystal can be described by an effective Hamiltonian (Supplements A and B) acting on a four-component wavefunction. In the more rigorous second approach, we construct the effective Hamiltonian starting with the trivial four-fold degeneracy at K-point and introduce two classes of perturbation: i) the dielectric perturbation induced by the removal of the circular segment of the disc leading to the bianisotropy, and ii) the perturbation induced by the change in the propagation direction (deviation from the K point). The effect of both of these perturbations was found from numerically calculated unperturbed field profiles (i.e. for the symmetric disk at the K point) by using equations of degenerate perturbation theory (Supplement A). The theory can be summarized by two equations which allow calculating matrix elements of the 4x4 Hamiltonian (up to a normalization factor):

$$\Delta_{mn} = -\int d^3r \, \delta\epsilon_r \, \epsilon_0 \mathbf{E}_n \cdot \mathbf{E}_m^*, \tag{1}$$

$$\{S_i\}_{mn} = \int_V d^3r \, \{\mathbf{E}_n^* \times \mathbf{H}_m + \mathbf{E}_m \times \mathbf{H}_n^*\}_i, \tag{2}$$

where $\Delta_{mn}$ describes the change in the Hamiltonian due to the perturbation of the dielectric constant by $\delta\epsilon_r(\mathbf{r})$, and $\{S_i\}_{mn}$ reflects the change due to the displacement of the Bloch vector from the K point, and $\mathbf{E}_n$ and $\mathbf{H}_n$ are electric and magnetic fields of the (unperturbed) n-th eigenmodes. This procedure allows us to restore the Hamiltonian in the form:

$$\widehat{\mathcal{H}}_K = \omega_0 + v_\parallel \hat{s}_0(\delta k_x \hat{\sigma}_x + \delta k_y \hat{\sigma}_y) + v_\perp \hat{s}_y \hat{\sigma}_z \delta k_z + m \hat{s}_z \hat{\sigma}_z, \tag{3}$$

where $\hat{s}_i$ and $\hat{\sigma}_i$ are Pauli matrices operating in the subspaces of polarization (pseudo-spin) and orbital momentum (dipole/p-orbital orientation), respectively, $m$ is the effective mass term induced by the bianisotropy, $\omega_0$ is the frequency of the Dirac bands at K-point, and $v_{\parallel(\perp)}$ is the in-plane (out-of-plane) Dirac velocity. The Hamiltonian Eq. (3) describes the 3D Dirac conical dispersion with the mass term inducing a band gap opening and leading to appearance of the surface state at the domain walls (Supplement C). Note that the Hamiltonian (3) also provides a local description near the K′-point with the sign of the mass term reversed $m \to -m$.

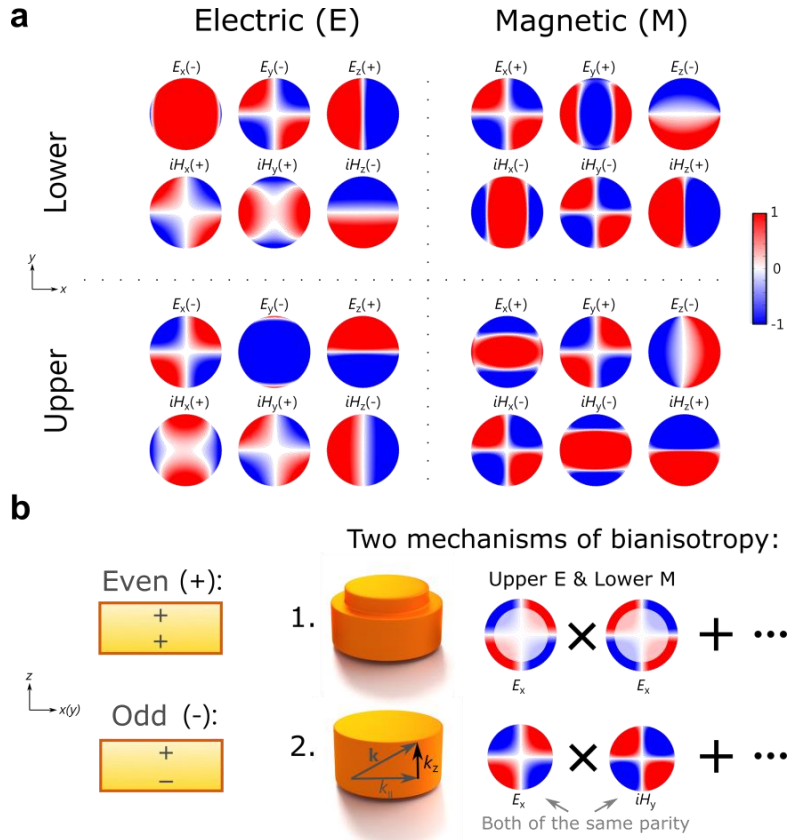

**Figure 2| Field profiles of the four bands corresponding to the overlaid 3D Dirac cones and the two ways to induce bianisotropy.** (a) Fields in the vicinity of the $K$ point for the structure shown in Fig. 1 before the perturbations – out-of-plane meta-atom reflection symmetry reduction and out-of-plane propagation ($k_z \neq 0$) – are introduced. The $x$-$y$ cut is near the base of the cylinder. (b) Left: +/- notation for even (electric) and odd (magnetic) modes. Right: two ways to induce bianisotropy and mix electric and magnetic modes of the cylinders by (1) their geometric symmetry reduction (only the unshaded regions contribute to Eq. (1)) or (2) out-of-plane propagation with the resultant mixing of electric and magnetic field components (as in Eq. (2)).

Interestingly, the structure of the Hamiltonian can also be understood from the field profiles of the modes shown in Fig 2(a). Specifically, the mass term is induced by a conventional bianisotropy, i.e. by coupling of even (electric) and odd (magnetic) modes, and, as it follows from Eq. (1), can be schematically explained as the mixing of the same electric field components in Fig. 2(b) (the first mechanism). The physics of the gap opening due to the spatial dispersion is distinct and, according to Eq. (2), it can be understood as the result of mixing of the electric and magnetic components of different eigenstates. It is instructive to apply this approach to explain the gap opening for finite values of $\delta k_z = k_z$, which leads to 3D Dirac dispersion in our system. In this case gap opening for symmetric cylinders appears due to the second mechanism of bianisotropy illustrated in Fig. 2(b) and caused by the mixing in the pseudo-spin subspace between electric and magnetic modes, which gives rise to the third term in the Hamiltonian Eq. (3). In the case of the in-plane propagation $\delta \mathbf{k}_{||} \neq 0$ (for symmetric cylinders and $\delta k_z = 0$), the mixing of electric and magnetic modes does not occur and gap opens solely due to the hybridization in the orbital subspace, which gives rise to the second term in Eq. (3).

The effective photonic Hamiltonian Eq. (3) is equivalent to a family of condensed matter Hamiltonians locally describing 3D topological insulators [47]. Therefore, the system under study represents a photonic equivalent of an electronic 3D topological insulator with the bi-anisotropy playing the role of the effective spin-orbit interaction. Note that similar equivalency of photonic and electronic topological systems has been established earlier for 2D bianisotropic metacrystals with TR symmetry made of metallic split-ring resonators [15] and a parallel plate waveguides with an array of metallic rods [31, 23]. Therefore, the proposed 3D all-dielectric system shares its topological classification with these 2D structures, and our 3D structure exhibits a weak $Z_2$ topological order common for stacked 2D topological systems [48, 49].

One of the most important consequences of topological order is the emergence of gapless surface states, which may occur either on the external interfaces with topologically trivial system or at the domain walls separating topological systems of distinct classification. While in condensed matter one typically considers a natural boundary of topological insulator with topologically trivial vacuum, in electromagnetics, such open boundary would necessarily lead to leakage of the surface states, at least for some subset of the wave-vectors lying above the light cone. However, even for modes lying below the light cone, any encounter of a defect (even a topology preserving defect) will result in scattering into free space. The other seemingly natural choice of perfect electric conductor (PEC) bounding the topological structure, while known to work in systems with TR symmetry broken by magnetization [8, 9], is also not always suitable for SPT systems as it may break the symmetry underlying the topological order, including ours. Indeed, since PEC boundary condition has a different effect on the electric and magnetic components of the fields it violates the duality, which results into a breakdown of the pseudo-spin degree of freedom and prevents formation of the topological surface state at the interface.

A two-dimensional topological domain wall separating two 3D structures with the opposite signs of effective mass $m$, i.e. the opposite bianisotropy, fulfils both requirements: (i) it preserves the duality and the pseudo-spin degree of freedom and (ii) prevents leakage of the surface states due to the presence of a complete 3D band gap in both of the topological half-spaces. From the topological classification of our system one can predict that some of such 2D domain walls should host topological surface states exhibiting linear 2D Dirac-like dispersion, with the Dirac cones that always appear in pairs at K and K′ points.

The analytical calculations within the effective Hamiltonian approach (Supplement C) confirm the presence of gapless surface states for vertical domain walls, e.g. $yz$ and $xz$ cuts, at both K and K' points. Within our analytical description given by Hamiltonian (3), a 2D domain wall with the flip of bianisotropy can be described as an interface across which the effective mass reverses its sign. In 2D systems, such as graphene and quantum spin Hall effect (QSHE) systems [50, 51, 7, 15], which are also described by the 2D Dirac equation, the mass term is known to open the band gap, and the reversal of the mass term across the interface gives rise to the emergence of the edge states. The system described by the effective Hamiltonian (1) displays analogous behaviour in 3D, and exhibits emergence of gapless surface states within the bandgap induced by the mass term, which was confirmed by finding the eigenmodes of Eq.(1) for two half-spaces and matching the solutions at the domain wall with the use of continuity boundary condition (Supplement C). However, unlike 2D Dirac systems described by two-component wave-functions, and in which the edge states exhibit linear 1D bands, the 3D system under study is locally (near the K and K' points) described by the four component wave-function, and exhibits 2D conical bands, thus even closer emulating relativistic Dirac equation. Indeed, the Hamiltonian (1) can be brought to a form identical to the Dirac equation by unitary linear transformation [47]. In this context, the surface modes can be viewed as the well-known Jackiw-Rebbi states which exist on the boundary between two half-spaces hosting relativistic quantum particles of opposite mass [46]. Thus, the system in hand can also be viewed as a classical emulator of 3D relativistic Dirac fermions.

To corroborate the analytical results, we performed a set of large scale full-vector numerical simulations with the use of finite element method (FEM) solver COMSOL Miltiphysics®. First, we consider the topological domain wall along the $yz$-cut in the middle of a supercell of 24 meta-atoms with the periodic boundary conditions imposed in all three directions. The domain wall represents an interface between two topological crystals whose meta-atoms have their narrow section facing opposite direction along the $z$-axis, which effectively results in the reversal of bianisotropy. The schematic view of the corresponding structure with the domain wall is shown in Fig. 3(a) with the band structure presented in Fig. 3(b,c). The two panels in Fig. 3(b) show two different cuts of the two-dimensional band structure $\omega(k_y, k_z)$ along high symmetry directions in the Brillouin zone and reveal two pairs of surface states. The surface states appear within the band gap region and cross it such that they interconnect lower and upper bulk bands. The 3D band diagram shown in Fig. 3(c) exposes one of the Dirac cones corresponding to the surface states near K-point (an identical cone at K' is not shown). The confinement of the surface states to the domain wall is confirmed by their field profiles for all directions within the domain wall plane. As an illustration, the corresponding electric field densities are plotted in Fig. 3(d), right panel, for several points on one of the isofrequency contours shown in Fig. 3(d), left panel. As in the case of 3D electronic topological insulators, the surface states in our system also have a property of spin-locking which endows the system with topological robustness. Both analytical results (Supplement C) and numerically calculated field profiles show that the pseudo-spin configuration is uniquely defined by the direction of the surface mode wave-vector $k$, as schematically indicated in Fig. 3(d), left panel, and the modes of the same pseudo-spin but opposite wave-vectors (lying on a second Dirac cone near the K′ point) are opposite to each other.

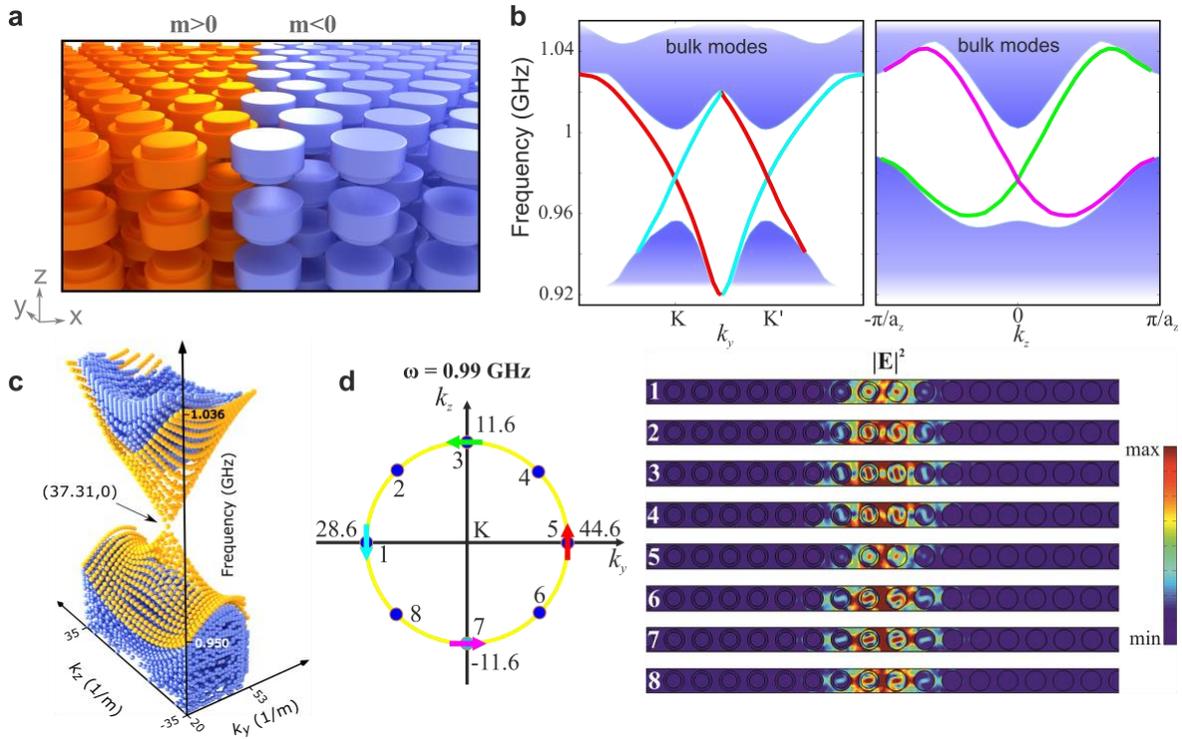

**Figure 3| Topological surface states of two-dimensional *yz*-domain wall in 3D all-dielectric metacrystal.** (a) The schematics of the domain wall formed by the reversal of the mass term induced by the bianisotropy in the middle of the metacrystal. (b) Band diagrams of topological surface states supported by the domain wall in (a) with two-dimensional cut-planes $\omega$-$k_y$ and $\omega$-$k_z$ shown in the left and right panels, respectively. (c) The conical Dirac-like dispersion of the surface states. (d) The isofrequency contour on the Dirac cone (left panel) and the field profiles (right panel) indicating localization of the surface state to the domain wall for different locations of the wave-vector on the cone.

The edge and surface states of topological systems are of significant interest due to their unique property of topological robustness that manifests as a reflectionless propagation and insensitivity to defects and disorder which do not violate a symmetry underlying a particular SPT order. Here we show that the surface states of the 3D all-dielectric system in hand exhibit a similar property. To this end, we performed the first-principles (full-wave) numerical studies of the surface states propagating along a curved topological domain-wall with two sharp bends, as shown in the top left panel of Fig. 4. In contrast to the edge modes confined to 1D interfaces of 2D topological systems, the 2D surface states of our 3D metacrystal can propagate in two dimensions. This allows testing their robustness for multiple values of the out-of-plane wave number $k_z$, which is equivalent to sweeping the wave-vector along the circular isofrequency contour in Fig. 3(d). The results of the numerical simulations for the vertical cut of the domain wall are shown in Fig. 4, and clearly demonstrate that the field intensity remains uniform as it propagates along the domain wall regardless of the bends. Thus, the wave avoids back-reflection from the sharp bends which, in particular, circumvents formation of the standing-wave interference patterns that would necessarily occur in the straight linear segments of any conventional non-topological waveguide [31, 21, 23]. In 2D photonic systems this robustness of the edge modes has been attributed to the property of locking of the pseudo-spin, a specific polarization configuration of the wave, to its propagation direction [15, 31, 21, 23]. The same holds true for the 3D system studied here with the exception that the pseudo-spin of the mode changes continuously as the wave-vector sweeps around the Dirac cone (Fig. 3c, Supplement C).

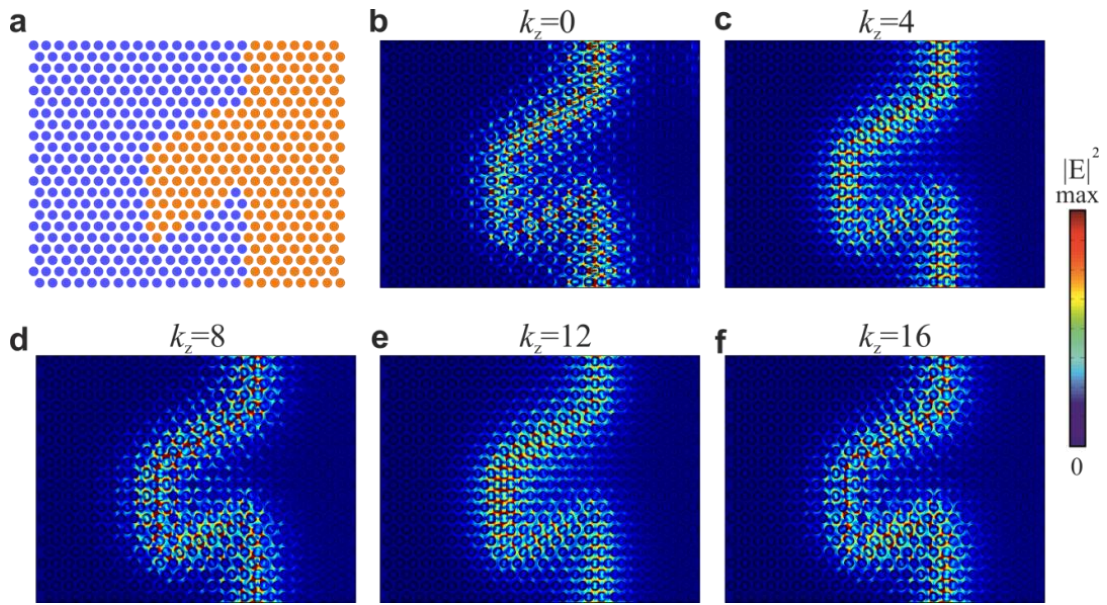

**Figure 4| Topological robustness of surface states propagating along a sharply curved 2D domain wall formed in the middle of the all-dielectric 3D topological insulator.** (a) The shape of the domain wall: the blue and orange circles indicate "up" and "down" orientations of the meta-atoms, respectively. (b)-(f) The field distribution of the surface modes propagating without reflection across the domain wall with the sequence of sharp bends as in (a). The different subplots correspond to different values of the out-of-plane wave-vector component $k_z$.

Another major distinction of the proposed 3D photonic structure from its 2D counterparts is related to the possibility of having domain walls facing any direction in 3D space. As an example, here we consider the domain wall in the vertical (along the *z*-axis) direction, which is formed by stacking the all-dielectric metacrystals so that upper and lower domains have wider sections of the discs facing each other, as shown in Fig. 5(a). The analytical description of this situation with the flip of the mass term in the vertical direction in the context of the effective Hamiltonian formalism Eq.(1) predicts existence of the surface states with conical 2D Dirac-like dispersion.

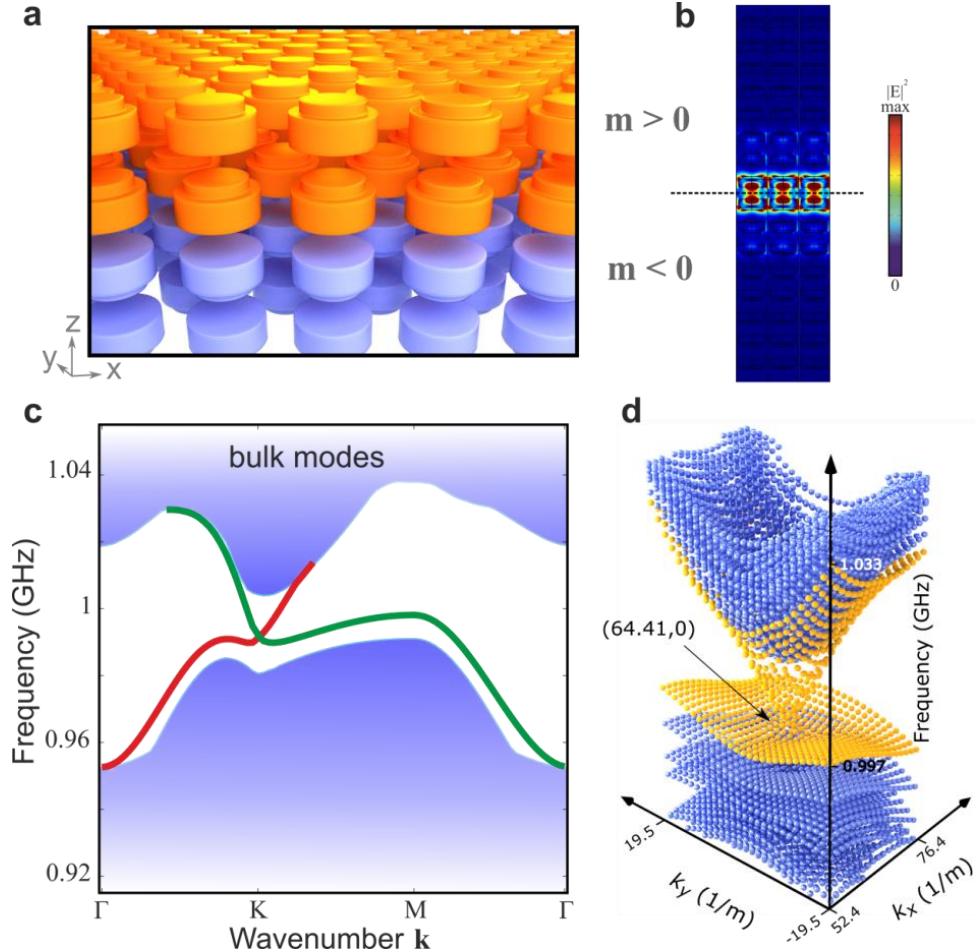

**Figure 5| Surface states of two-dimensional *xy*-domain wall in three-dimensional all-dielectric metacrystal.** (a) The schematics of the domain wall formed by the reversal of the bianisotropy in the middle. (b) The field profile indicating localization of the surface state to the domain wall at K point. (c,d) Band diagrams of topological surface states supported by the domain wall in (a). (c) Two-dimensional band diagram $\omega$ vs. $\mathbf{k}_\parallel$, with $\mathbf{k}_\parallel$ changing along the high symmetry directions of the 2D hexagonal Brillouin zone of the domain wall, and (d) the conical Dirac-like dispersion of the surface states near the K-point of the 2D hexagonal Brillouin zone.

The numerical simulations of the all-dielectric metacrystal confirm these predictions with the corresponding field profile of the surface state shown in Fig. 5(b) and photonic band diagrams plotted in Fig. 5(c,d). Note that the corresponding (*xy*) domain wall has hexagonal symmetry and effectively represents a quasi-2D lattice with the modes confined in the vertical direction due to the presence of the complete band gap of upper and lower metacrystals. We found that, in general, the band structure of the domain wall reveals two types of bands within the band gap region. The first type of bands represents the surface modes extending all the way through the band gap. In addition, the structure hosts the second type of modes which also tend to localize on the wall, but do not cross the bandgap. The latter type of modes represents conventional Tamm-like defect modes localized near the domain wall and are not shown here. Note that, in general, the domain wall may host a number of such states, which, depending on the staking configuration, can spectrally overlap with the surface states of the first type. Nonetheless, the Tamm states can always be removed from the bandgap region by adiabatically changing the configuration of the stacking, which, in particular case of structure shown in Fig.5, was achieved by tuning the separation between upper and lower metacrystals. The resultant isolated Dirac-like conical bands, demonstrating the presence of standalone surface states within the entire gapped region, are shown in Fig. 5(d). As opposed to Tamm states, these surface modes, cannot be removed by any deformation preserving the in-plane symmetry of the domain wall. However, as one could expect for the weak topological structure formed by stacking planar QSHE layers, these modes are topologically trivial by their nature. They exist solely due to the domain wall playing the role of a high index defect, which facilitates localization of the field, and the in-plane hexagonal symmetry, which ensures the presence of the Dirac points at the K and K′ valleys. Nonetheless, it is indeed a noteworthy property of our system that while representing a weak photonic topological insulator it may exhibit gapless surface modes along both horizontal and vertical cuts.

While in the design above we have used meta-atoms with high dielectric permittivity $\epsilon_d = 81$ embedded into the matrix with $\epsilon_b = 3$. The only purpose of high dielectric contrast was to allow strong localization of the surface modes, which reduces the size of the geometrical domain used in numerical modelling. Nonetheless, in the microwave frequency range materials with high permittivity and low loss are easy to find. For example, the dielectric constant for water is $\epsilon = 81$, for ceramic MgO-CaO-TiO$_2$ it ranges from $\epsilon \approx 20$ to $\epsilon \approx 140$, and for ceramic TiO$_2$-ZrO$_2$ lies within the range $95 - 100$, depending on the composition. Any type of plastics can be involved in order to realize the matrix, for example polyethylene, to host the cylinders. In the optical domain, the realization of the structure is more challenging, but still possible with the use of high-index materials, Ge$_2$Sb$_2$Te$_5$ $\epsilon \approx$30 [52] and tellurium (Te) $\epsilon \approx$23 in infrared [53], and silicon (Si) $\epsilon \approx 12$ and germanium (Ge) $\epsilon \approx 16$ in near-infrared, which can be patterned by 3D nanofabrication techniques such as NanoScribe®. We have found that to implement a practical design for optical domain the lack of a very high permittivity can be compensated by reduction of the matrix permittivity. And although the surface states tend to be less localized to the domain walls due to the reduction in the width of the topological bandgap, they remain to be well defined.

In conclusion, we have demonstrated that the all-dielectric metacrystal proposed here represents a classical toy model emulating relativistic quantum fermions that offers unique opportunities to study the fundamental physics in predictable and highly controllable manner. Such handmade table-top all-dielectric photonic structures can be precisely engineered to exhibit fascinating physical phenomena which so far evaded observation. In particular, the observation of the surface states predicted here would provide the first experimental evidence on existence of exotic Jackiw-Rebbi excitations. Besides these fundamental aspects, the topological robustness of the surface states enables reflectionless routing of electromagnetic radiation along arbitrarily shaped pathways in three dimensions, which makes such modes promising for applications in photonics. The proposed concept envisions implementation of topologically robust three-dimensional photonic circuitry in entire electromagnetic spectrum from microwave frequencies to optical domain. This all-dielectric platform will allow avoiding undesirable effects of Ohmic loss necessarily present in metallic and plasmonic structures. It facilitates topological order in photonics-compatible dielectric and semiconductor materials, thus escaping technical difficulties associated with the use of magnetic materials and external magnets required to induce topological order in systems with broken TR symmetry.

**Acknowledgments**

AK acknowledges support by the National Science Foundation (CMMI-1537294). Research was carried out in part at the Center for Functional Nanomaterials, Brookhaven National Laboratory, which is supported by the U.S. Department of Energy, Office of Basic Energy Sciences, under Contract No. DE-SC0012704.

**Author contributions**

All authors contributed extensively to the work presented in this paper.

**Additional information**

Correspondence and requests for materials should be addressed to A.K.

**Competing financial interests**

The authors declare no competing financial interests.